\documentclass[pre,superscriptaddress,twocolumn,final,showpacs,showkeys,nobibnotes,titlepage,twoside,10pt]{revtex4-1}
\usepackage{amsmath}
\usepackage{graphicx}
\usepackage{amsfonts}
\usepackage{amssymb}
\usepackage{subfigure}
\usepackage{float}
\usepackage{color,soul}
\usepackage{natbib}
\begin{document}

\author{A. E. Lagogianni}
\affiliation{I. Physikalisches Institut, Universit\"at G\"ottingen, Friedrich-Hund-Platz 1, 37077 G\"ottingen, Germany}
\author{J. Krausser}
\affiliation{Statistical Physics Group, Department of Chemical Engineering and Biotechnology,
		University of Cambridge, New Museums Site, Pembroke Street, CB2 3RA Cambridge, U.K.}
\author{Z. Evenson}
\affiliation{Heinz Maier-Leibnitz Zentrum (MLZ) and Physik Department, Technical University Munich, Lichtenbergstrasse 1,
85748 Garching, Germany}
\author{K. Samwer}
\affiliation{I. Physikalisches Institut, Universit\"at G\"ottingen, Friedrich-Hund-Platz 1, 37077 G\"ottingen, Germany}
\author{A. Zaccone}
\affiliation{Statistical Physics Group, Department of Chemical Engineering and Biotechnology,
		University of Cambridge, New Museums Site, Pembroke Street, CB2 3RA Cambridge, U.K.}
\affiliation{Cavendish Laboratory, University of Cambridge, CB3 0HE Cambridge, U.K.}
\date{\today}

\begin{abstract}
An analytical framework is proposed to describe the elasticity, viscosity and fragility of metallic glasses in relation to their atomic-level structure and the effective interatomic interaction. The bottom-up approach starts with forming an effective Ashcroft-Born-Mayer interatomic potential based on Boltzmann inversion of the radial distribution function $g(r)$ and on fitting the short-range part of $g(r)$ by means of a simple power-law approximation. The power exponent $\lambda$ represents a global repulsion steepness parameter. A scaling relation between atomic connectivity and packing fraction $Z \sim \phi^{1+\lambda}$ is derived. This relation is then implemented in a lattice-dynamical model for the high-frequency shear modulus where the attractive anharmonic part of the effective interaction is taken into account through the thermal expansion coefficient which maps the $\phi$-dependence into a $T$-dependence. The shear modulus as a function of temperature calculated in this way is then used within the cooperative shear model of the glass transition to yield the viscosity of the supercooled melt as a double-exponential function of $T$ across the entire Angell plot. The model, which has only one adjustable parameter (the characteristic atomic volume for high-frequency cage deformation) is tested against new experimental data of ZrCu alloys and provides an excellent one-parameter description of the viscosity down to the glass transition temperature.

\end{abstract}

\title{Unifying interatomic potential, $g(r)$, elasticity, viscosity, and fragility of metallic glasses: analytical model, simulations, and experiments}
\maketitle

\section{Introduction}
\subsection{State of the art}
One of the most puzzling properties of glasses is the huge increase of viscosity, by many orders of magnitude, within a narrow range of temperature $T$ upon approaching the glass transition temperature $T_{g}$. As a consequence, considerable interest is being devoted to understanding this phenomenon in terms of the underlying atomic-level structure and dynamics of supercooled liquids. In recent years, much attention has been devoted to the roles of both short and medium-range order in promoting structural and dynamical arrest upon approaching the glass transition from the liquid side. Various order parameters have been proposed to quantify the local order in supercooled liquids, starting from the bond-orientational order parameter \cite{Royall2008}, until more recent proposals which showed evidence that glassy properties correlate strongly with the local breaking of inversion-symmetry at the atomic scale. As a matter of fact, the local inversion-symmetry breaking turns out to be the key microstructural aspect which controls both the nonaffine soft elasticity and the boson peak of glasses \cite{Milkus2016}. 

In terms of materials applications, there is little doubt that disordered glassy materials represent part of the future of materials science, due to their advanced applications, in particular in terms of their outstanding performance under mechanical loading \cite{Ketov2015}. Metallic glasses have emerged as the most important class of glassy materials from this point of view, and it remains one of the major challenges of current research to understand the relationship between atomic dynamics and macroscopic mechanical response in these materials. 

Due to the current limitations in terms of experimental and computational techniques, the most abundant and reliable information about the atomic-scale structure of liquid and glassy metals comes from two-point correlation functions such as the structure factor $S(q)$, which can be extracted from neutron and X-ray scattering, and gives access to the radial distribution function, $g(r)$. In the theory literature~\cite{Berthier}, attention has been devoted, over the last decades, to multi-point correlation functions such as the four-point correlation function $\chi_{4}$, which exhibits more significant changes upon crossing the $T_{g}$, whereas the $g(r)$ remains substantially unaltered upon going from the liquid into the glass.
In reality, appreciable changes in $g(r)$ can be detected upon vitrification, although the extent of these changes appears to vary from system to system, and this represents a possible way of linking structural evolution to dynamics \cite{Mauro2014}. 

Among the most popular pictures proposed to link the phenomenology of the Angell plot for the viscosity $\eta$ versus $T$ near $T_{g}$, is the one which associates fragile glass formers (with the steepest dependence of $\eta$ on $T$) to an underlying steep interparticle repulsion at contact, whereas strong glasses (with Arrhenius dependence of $\eta$ on $T$) are associated with softer interparticle repulsion. This picture, which is largely based on the two-point correlation dynamics and local structure, and on the Weeks-Chandler-Anderson \cite{Weeks1971} idea that the repulsive part of two-body interaction is what controls the overall structure of liquids, has been demonstrated convincingly for the case of soft colloidal glasses by the Weitz group \cite{Mattson2009}. 

\subsection{"Soft atoms make strong glasses"}
Using a theoretical argument based on the high-frequency quasi-affine shear modulus and its relation to interatomic connectivity and thermal expansion, some of us~\cite{Krausser} recently showed that this picture ("soft atoms make strong glasses") may be applicable to metallic glasses as well. In particular, we showed that a global interatomic repulsion parameter can be defined from theory and can be linked to the ascending part of $g(r)$ to describe the full Angell plot (from strong to fragile) for metallic alloys of very different composition. In this theory, an important role is played also by thermal expansion: the fragile behaviour is linked with higher values of the thermal expansion coefficient $\alpha_{T}$. This is another "global" parameter, this time related to the longer-ranged anharmonic part of the interaction, which may also account, in a coarse-grained way, for effects of medium-range order. Both these global interaction parameters, $\lambda$ and $\alpha_{T}$, are sensitive functions of the elemental composition and stoichiometry of the alloy, and may account for microalloying effects as well \cite{Huang2011}.

\subsection{Medium-range atomic dynamics}
Other very recent studies have appeared which substantially support this picture from different angles. Busch and co-workers \cite{Wei2015} have shown, for various alloys, that fragile behaviour indeed correlates with higher thermal dilation, and with shallow changes in the first coordination shell. The first feature, is seen in the structural evolution on a scale of 3-4 atomic diameters, where long-range interaction effects and anharmonicity control the incorporation of volume upon changing $T$. The shallow structural change observed at the level of first coordination shell for fragile melts, instead, can be explained by the steepness of the potential: upon reducing $T$, systems with steeper repulsion experience an increased resistance towards further approach of two nearest-neighbours, thus leading to the observed shallow variation in the position of the first maximum of $g(r)$. 
In recent work, this picture has also been connected to the Arrhenius crossover temperature, which marks the appearance of the high-T liquid regime where full Arrhenius behaviour of transport properties is recovered \cite{Jaiswal,Wang2015}. 

\subsection{5-fold symmetry picture}
In a somewhat different picture proposed recently by Wang and co-workers \cite{Hu2015}, fragility was found to correlate with the rapidity at which 5-fold (icosahedral-like) symmetry 
develops upon lowering $T$ \cite{Lagogianni2009,Bokas2013}. Although apparently a different picture, this finding can be related to the interatomic softness parameter $\lambda$ of Ref.{\cite{Krausser}. Atoms tend to organize themselves in icosahedral clusters with their nearest-neighbours, as $T$ decreases \cite{Almyras2011,Antonowicz2012} . This process becomes faster with steeper interatomic repulsion (larger $\lambda$) because the re-organization energy, when two atoms are not too close to each other, is comparatively less than for softer potentials where at the same distance atoms still experience significant energy. 

\subsection{Analytical relations for $\eta(T)$}
Finally, several of these recent works sought an analytical relationship for the increase of viscosity $\eta$ as a function of $T$ in the Angell plot. Kelton and co-workers~\cite{Mauro2014} provided an extended Vogel-Tammann-Fulcher (VTF) relationship which involves a structural parameter related to the change in the first coordination shell. Wang and co-workers \cite{Hu2015} proposed another extension of the VTF relationship which, instead, involves an order parameter for 5-fold symmetry. Both these relations have two free fitting parameters (in addition to the normalization constant $\eta_{0}$).

An altogether different relationship, with just one adjustable parameter, has been proposed by Johnson and co-workers~\cite{Johnson2007} and further developed in Ref.\cite{Krausser}.
This relation, furthermore, is not of the VTF type, but arises from a microscopic mechanistic picture given by the shoving model of the glass transition due to Dyre, made microscopic by using an atomic theory for the high-frequency shear modulus involving the interatomic repulsion parameter $\lambda$ and the thermal expansion coefficient $\alpha_{T}$. The only adjustable parameter is the characteristic atomic volume $V_{c}$ for cage deformation, which was found to be related, on a larger scale, to the volume of shear-transformation zones (STZ) \cite{Jensen2014}. 

In this contribution, we apply theoretical ideas~\cite{Johnson2007,Krausser} to experimental data of metallic glasses to show that a bottom-up quantitative relationship can be built between the atomic-scale structure and interactions and the macroscopic $T$-dependent viscosity. The resulting framework, in which a major role is played by globally-averaged interaction parameters related to short-ranged interatomic softness and longer-ranged anharmonicity, uses the short-range $g(r)$ as input from simulation or experimental data, to arrive at the viscosity for which the theory provides an excellent one-parameter fitting.
This good agreement shows that changes at the level of first-coordination shell, encoded in the short-range repulsion parameter $\lambda$, as well as changes at the medium-range scale, encoded in the thermal expansion coefficient $\alpha_{T}$, are both important in determining the viscosity and fragility of metallic glasses.

\section{Interatomic potential for the ion-ion repulsion in metallic glasses}
In recent work~\cite{Krausser}, we analysed several alloys in an attempt to extract an effective, averaged interatomic potential which describes the short-range repulsion between any two ions in a metallic alloy melt. Based on the systematic fitting of shear modulus and viscosity data for various three- and 5-component alloys we proposed the following interatomic potential which comprises two contributions: (i) the longer-ranged Thomas-Fermi screened-Coulomb repulsion modulated by the Ashcroft correction and (ii) the Born-Mayer closed-shell repulsion due essentially to Pauli repulsion. The Thomas-Fermi contribution is more long-ranged and is described by a Yukawa-potential type expression. The Born-Mayer contribution is a simple exponentially-decaying function of the core-core separation, motivated by the radial decay of electron wavefunctions for the closed shells. The effective interatomic potential reads as
\begin{equation}
V(r)=A \frac{\exp^{-q_{\text{TF}}(r-2a_0)}}{r-2a_0}+B\mathrm{e}^{ -C(r-\bar{\sigma})}
\end{equation}
where 
\begin{equation}
A=Z_{\text{ion}}^2 e^2\cosh^{2}(q_{\text{TF}}  R_{\text{core}})
\end{equation}
is the Ashcroft factor~\cite{Faber1972}, with $ R_{\text{core}}$ being a typical value for the atom-specific core radius and  $Z_{\text{ion}}$ the effective ionic charge number.
Furthermore, $a_0$ is the Bohr radius and $\bar{\sigma}$ is the average ionic core diameter of the alloy, which corresponds to the average size of the ionized atoms constituting the alloy. The average ionic core diameter is obtained by averaging the respective ionic core diameter of the constituents with their contributing weights given by their volume ratios in the alloy. The values for the ionic core diameters of the atoms constituting the alloys are taken from Ref.~\cite{Shannon1976}. 

The quantities $A$ and $B$ set the energy scales for the repulsive interaction from the Ashcroft and Born-Mayer term, respectively.
The parameter $q_{\text{TF}}$ is the inverse of the Thomas-Fermi screening length given by Thomas-Fermi theory, and its value is known for different types of alloys~\cite{Wang2004}. We choose a representative value for $q_{\text{TF}}$ as $1.7$~$\mathrm{\AA^{-1}}$ according to the values reported in Ref.~\cite{Wang2004}. The ionic core diameter $\bar{\sigma}$ is obtained by a weighted average of the core diameters of the atoms constituting the alloys taken from~\cite{Shannon1976}, where the weights correspond to the ratios of the respective atoms. 

The characteristic range $1/C$ of the valence-shell overlap repulsion is not known \textit{a priori}. However, its typical values are less sensitive to the atomic composition than the parameters $\bar{\sigma}$, $A$ and $B$. Different atoms have very similar values typically in the range $C=1.89 - 4.72$~$\mathring{A}^{-1}$~\cite{Nikulin1970}. 

The latter cannot be easily estimated from first-principles or from literature. Similarly, the prefactor $B$ of the Born-Mayer term, can be rigorously evaluated only from the exchange integrals of the various overlapping electrons belonging to the valence shells of the two interacting ions. This calculation, even in approximate form, is not feasible except for simple monoatomic crystals. Hence, both $A$ and $B$ were taken in our previous analysis as adjustable parameters in the mapping between our schematic logarithmic potential (to be introduced below in Sec.III) and the Ashcroft-Born-Mayer interatomic potential. We shall remark that the Born-Mayer prefactor $B$ typically has non-trivial large variations from element to element across the periodic table, as shown in many ab initio simulation studies~\cite{Nikulin1970, Hafner1987}. Consistent with this known fact, it turns out that $B$ is the most sensitive parameter in our analysis, in the sense that small variations in $B$ can lead to large deviations in the fitting of the experimental data. Conversely, the Ashcroft prefactor $A$ is not a sensitive parameter, and its values are not crucial for the match with experiments. 

In Ref.~\cite{Krausser}, it was found that, in order to fit shear modulus and viscosity data of various alloys, values of the Born-Mayer repulsion strength $B$ are required which are between two and three orders of magnitude smaller than the Born-Mayer parameters tabulated for pure metals. This important difference has at least two reasons. One reason is that the Born-Mayer formula used for pure substances in the literature is written $\sim \exp{[-Cr]}$, instead of $\sim \exp{[ -C(r-\bar{\sigma})]}$, which we use here. Evidently, $\exp{[C\bar{\sigma}]}>1$ partly contributes to explain this discrepancy. However, the fact that $B$ fitted for multi-component alloys is much smaller than $B$ found for pure substances is also due to the so-called micro-alloying effect, whereby the addition of even a small amount of different elements with different ionic size and electronic structure induces a strongly nonlinear change in the inter-ionic potential. When there is an atomic size mismatch, this difference intuitively promotes softer repulsion due to the fact that short-ranged packing is more efficient. Of course there could be other important reasons related to the change in electronic structure, e.g. the change in anisotropy of closed electronic shell, which also make the effective short-range repulsion in alloys being effectively milder than in pure metals. Finally, the Born-Mayer parameters in Ref.~\cite{Abrahamson1969} refer to pairs of atoms, in which the outer electronic structure is clearly much different from metals, where the atoms are significantly ionized. 

\section{The global interatomic repulsion parameter $\lambda$}
In the analysis of Ref.~\cite{Krausser}, it was found that different disordered alloys can be described by an effective interatomic potential given by Eq.(1) with $A=0.3-0.5~\mathrm{eV}$, and $B=5-60$~$\mathrm{eV}$. The value of $\bar{\sigma}$ is uniquely determined by the elemental composition of the alloy, whereas $q_{TF}=1.7$~$\mathrm{\AA^{-1}}$ can be kept constant, independent of composition. 
Furthermore, an even simpler parametrization of the interatomic potential was obtained by mapping Eq.(1) onto a logarithmic expression with a single, global parameter 
$\lambda$, which contains all the information about the steepness (or its inverse, the softness) of the ion-ion repulsion. This simpler expression reads as
\begin{equation}
V(r)=-\lambda \ln(r-2a_{0}).
\end{equation}

The mapping between Eq.(1) and Eq.(3) is shown in Fig.1, for representative parameters of metallic glasses. In all instances examined thus far, Eq.(1) is very accurately represented by Eq.(3). 

\begin{figure}
\centering
\includegraphics[width=0.94\columnwidth]{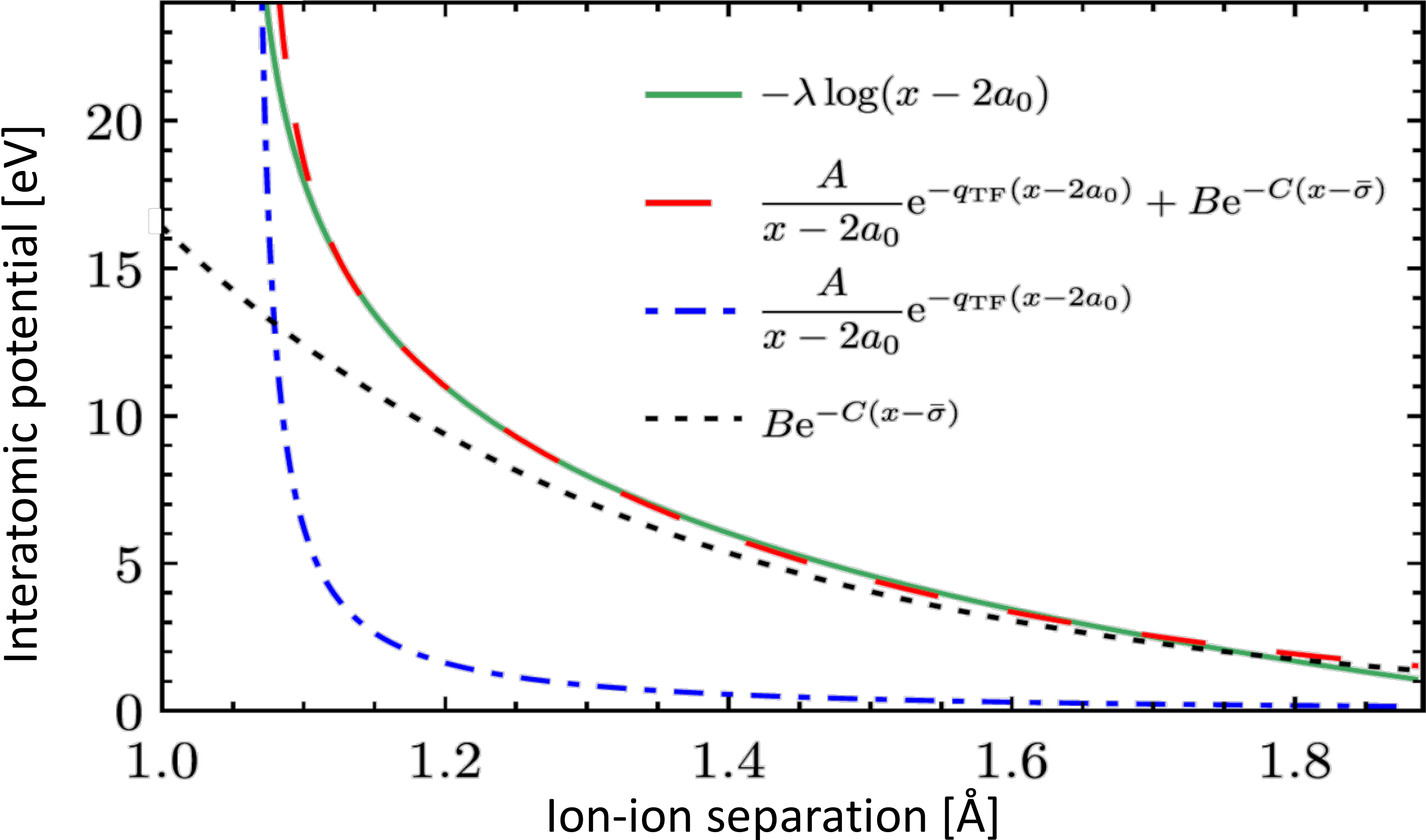}
\caption{(Color online)
Representation of the Ashcroft–Born–Mayer interatomic potential Eq.(1) using the one-parameter logarithmic expression Eq.(3) in terms of the global interaction parameter $\lambda$ (including the two separate contributions to the interatomic potential). This illustrative plot was generated for a repulsive steepness $\lambda =99.7$. 
}
\label{fig:rn}
\end{figure}

The main advantage of representing the repulsive part of the interatomic potential) with the compact Eq.(3) is that it allows us to pack various effects into a single, global interaction parameter $\lambda$ which contains information about the overwhelmingly complex details of the ion-ion interaction in metallic alloys. The interaction between two ions in metallic glasses is in fact the result of the intricate underlying electronic structure as well as of many-body and non-local effects. It is a hopeless task to devise a theory of the interatomic interaction due to this complexity, and our aim here is to present an averaged non-local parameter which, similar to the effective mass concept in semiconductors, takes all these non-trivial effects into account while still allowing one to label different alloys and their properties in terms of the microscopic interaction. 

A second important advantage of the compact form, Eq.(3), is that it allows us to relate the repulsion parameter $\lambda$ directly to the short-range ascending slope of the radial distribution function $g(r)$. Upon identifying the effective interatomic potential with the potential of mean force, Boltzmann inversion provides a link between the effective potential of mean force between two ions and the local structure
\begin{equation}
V(r)/k_{B}T=-\ln g(r).
\end{equation}

Importantly, Boltzmann inversion~\cite{Hafner1987} provides a definition of $V(r)$ as a non-local interaction potential which contains important many-body contributions, since the potential of mean force defines the effective interaction of two ions in the field of all the other ions and many-body interactions thereof. 
Upon combining Eq.(4) with Eq.(3), we obtain a relationship between $\lambda $ and $g(r)$, in the form of the following simple power-law expression
\begin{equation}
g(r)\sim(r-\sigma)^{\lambda}
\end{equation}

The link between the $g(r)$ and the interatomic potential $V(r)$ is schematically illustrated in Fig. 2.

\begin{figure}
\includegraphics[width=0.94\columnwidth]{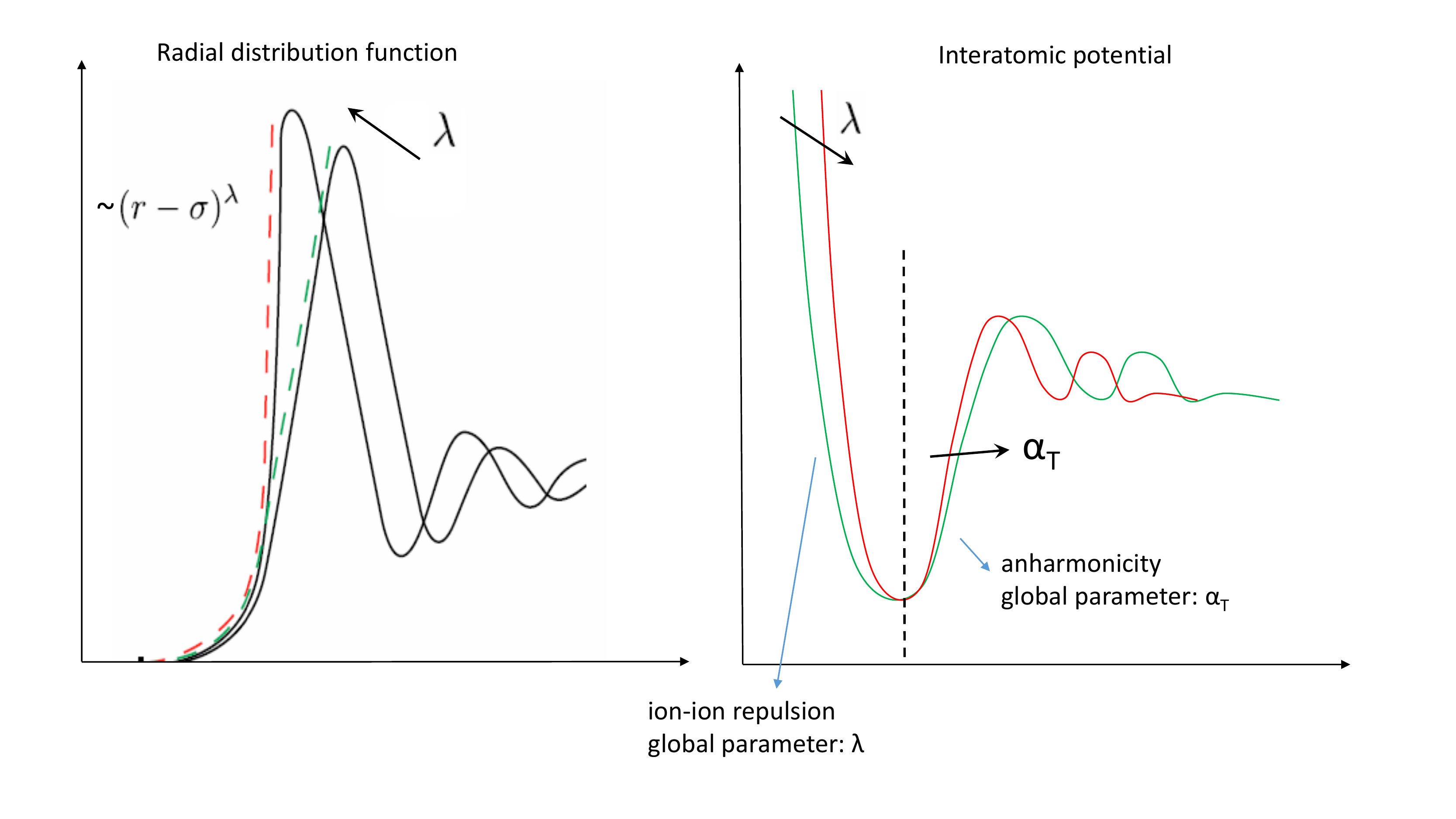}
\caption{(Color online)
The global repulsion parameter $\lambda$ and its relation to the atomic-scale structure. The radial distribution function $ g(r) $ (left panel) is related to the interatomic potential) $ V(r) $ (right panel) via Eq. (4). While the parameter $\lambda$ models the effective repulsion between two ions mediated by the field of all other ions in the material, the attractive part of the potential of mean force can be parameterized via the Debye-Grueneisen thermal expansion coefficient $\alpha_{T}$. 
}
\label{fig2}
\end{figure}

\section{Analytical expression for the high-frequency shear modulus}


In this section we develop a link between $\lambda$ and the high-frequency shear modulus. As is known at least since the seminal work of Zwanzig and Mountain~\cite{Zwanzig1965}, the high-frequency shear modulus of liquids can be efficiently described using affine or quasi-affine elasticity. In brief, the atoms forming the transient "cage" around a given atom are simply displaced proportionally to the imposed strain, with the high-frequency of the external driving not allowing for nonaffine rearrangements. The latter dominate instead the low-frequency response of liquids and glasses and are responsible for a dramatic softening of G, as they are associated with a negative contribution to the elastic free energy (in fact their contribution is internal work done by the system, hence negative by thermodynamic principles)~\cite{Lemaitre2006,Zaccone2011a,Zaccone2011b,Zaccone2013, Zaccone2014,Rizzi2016}.

Importantly, nonaffine response theory of Ref.~\cite{Lemaitre2006,Zaccone2011a,Zaccone2013} does correctly recover the Maxwell marginal rigidity criterion at the isostatic point at which the total number of constraints $ZN/2$ (where $Z$ is the mean number of bonds per atom) is exactly equal to the total number of degrees of freedom per atom $dN$, with central-force interactions (leading to $Z=2d=6$ at the isostatic point).
Conversely, previous approaches based uniquely on isostaticity~\cite{Wyart} and ignoring the symmetry, are less general and of limited applicability. For example, unlike the more general nonaffine formalism, isostaticity-based approaches  such as the one of Wyart~\cite{Wyart} provide erroneous predictions of the moduli of centrosymmetric lattices; for instance they predict the shear modulus of perfect centrosymmetric lattices to scale as $G\sim (Z-6)$ and not as $G\sim Z$ which is the correct scaling in view of the affine response ensured by local inversion-symmetry being enforced in centrosymmetric lattices. 

Hence, upon taking the infinite-frequency limit of the frequency-dependent nonaffine shear modulus from nonaffine response theory~\cite{Lemaitre2006,Zaccone2011a}, as shown in Ref.~\cite{Krausser} and numerically confirmed in Ref.~\cite{Rizzi2016}, the affine modulus is retrieved, which scales as $G\sim Z$, where $Z$ is the interatomic connectivity.
 
The link between the effective repulsion, encoded in $\lambda$, and the high-frequency $G$ goes by the way of the atomic connectivity $Z$, which defines the average number of nearest-neighbours around a test atom and can be estimated by counting all neighbours contained within the first peak of $g(r)$. This is a conservative way of counting nearest-neighbours, as opposed to e.g. also considering atoms beyond the maximum of $g(r)$, since these effectively experience a much reduced restoring force. Furthermore, these farther apart neighbours are within the presumably attractive part of the potential of mean force and the physics of this region is already accounted for in our model by the thermal expansion coefficient $ \alpha_T $.
Therefore, at frequencies much larger than the inverse Maxwell relaxation time of the liquid, $\omega \tau_{M} \gg 1$, the shear modulus can be evaluated using affine elasticity theory~\cite{Zaccone2011a}, which gives

\begin{align}
G = \frac{1}{5 \pi} \frac{\kappa}{R_0} \phi Z. 
\label{eq:6}
\end{align}

Here, $ \kappa $ stands for the spring constant of a harmonic bond and $ R_0 $ for the average interatomic spacing at rest in the equilibrated glass. The coordination number $Z$ refers to the average number of \textit{mechanically-active} nearest-neighbours~\cite{Zaccone2011a,Zaccone2011b,Zaccone2014}. While coordination numbers are typically defined from an integral over the first peak of $g(r)$, there is no consensus about the upper limit of the integral and on how this relates to the mechanical activity of the nearest-neighbours being counted.
For example, it is well known that, if the first minimum of $g(r)$ is taken as the upper limit of integration, then the integral yields $Z\approx 12$ both in the liquid and in the glass, and this result is basically independent of the temperature of the system. Clearly, this definition is totally inadequate to estimate the average number of mechanical contacts which is required by the expression for $G$.

The value of $g(r_{\mathit{max}})$ evaluated at the first peak position $ r_{max}$, increases significantly with increasing the packing fraction $\phi$ (and with decreasing $T$), which is typical of all dense liquids with an important repulsive component of the interaction~\cite{Egami2002}. This is related to the fact that nearest-neighbours located within the attractive minimum of $ V(r) $ are more likely to contribute to the rigidity.
Metals definitely fall into this category, and their $g(r)$ shares many features with the hard-sphere system, so that, for example, their $g(r)$ can be expressed in terms of the hard-sphere $g(r)$ using the celebrated Weeks-Chandler-Anderson method.  

Most importantly, the increase of the value of $g(r_{\mathit{max}})$ with increasing atomic density is modulated by the steepness of the interatomic repulsion, since the slope of the ascending flank of the first peak of $g(r)$ depends on the slope of the repulsive part of $ V(r)$ via the Boltzmann formula in Eq. (4). 
This fact can be intuitively understood by considering that the integral of $g(r)$ up to the maximum $g(r_{\mathit{max}})$ must necessarily yield a larger number when the slope of the ascending part of $g(r)$ is less steep compared to the case of a steeper slope (for the same maximum height). Hence, $Z$ must be an increasing function of the atomic density modulated by $\lambda$. 

Using the fact that the upper integration limit of $ r_{max} $ increases with the packing fraction $\phi$, integrating Eq.(5) up to a threshold which is proportional to $\phi$, as done in Appendix A, yields the scaling law $Z \sim  \phi ^{1+\lambda}$. Although the upper limit of the integral could be perhaps identified with $r_{\mathit{max}}$, since we are interested here in the qualitative behaviour we prefer to leave it as a generic threshold $\propto\phi$ such that the limit $Z\rightarrow 0$ is correctly recovered when $\phi \rightarrow 0$.

Moreover, the definition of the Debye-Gruneisen thermal expansion coefficient $\alpha_{T}$, in terms of the atomic packing fraction $\phi = vN/V$ (with $v$ the characteristic atomic volume and $N$ the total number of ions in the material) gives $\phi(T) \sim \mathrm{e}^{-\alpha_T T}$, as discussed in Ref.~\cite{Zaccone2013}. According to this result, $\phi$ decreases with increasing temperature $T$, an effect mediated by the thermal expansion coefficient defined as $\alpha_T = \frac{1}{V}(\partial V/ \partial T) = -\frac{1}{\phi}( \partial \phi/ \partial T)$.

Replacing the latter relationship between $\phi$ and $T$ in the equation for $Z$, see Eq. \ref{eq:6}, we finally obtain a closed-form equation which relates $ G $ to the two global interaction parameters, the short-range repulsion parameter $\lambda$ and the attraction anharmonicity parameter $\alpha_{T}$, 
\begin{equation}
G(T) =\dfrac{1}{5 \pi} \dfrac{\kappa}{R_0}\exp[ -(2+\lambda) \alpha_T T ].
\end{equation}
An exponential decay of the shear modulus with $T$ is found also in approaches like Granato's intersticialcy theory which model the glass as a crystal with a high concentration of interstitials, see e.g. Ref.~\cite{Khonik}.

Equation (7) accounts, in compact form, for all the salient features of the interatomic interaction, and contains the effect of repulsion steepness (short-ranged part of $V(r)$) as expressed by $\lambda$, and of anharmonicity, expressed by $\alpha_{T}$. A schematic depiction of how the global parameters $\lambda$ and $\alpha_{T}$ are related to features of $V(r)$, is presented in Fig. 2. It is important to note that the longer-ranged, anharmonic attractive part of the interaction in metals also stems from non-local, volume-dependent terms in the interaction of the ions with the partly delocalized electron gas. 
Hence, a microscopic description in terms of pair-interactions alone is generally not valid, although for volume-preserving shear deformations, as considered here, it can still be used.

The above expression, Eq.(7), can be rewritten as
\begin{align}\label{eq_th_shear}
G(T)=C_G\exp{\left[	\alpha_T T_g (2+\lambda)\left(1-\frac{T}{T_g}\right)\right]},
\end{align}
where $C_G=\frac{\varepsilon}{5\pi}	\frac{\kappa}{R_0}\mathrm{e}^{-\alpha_T T_g (2+\lambda)} $ is defined as the shear modulus value at the glass transition temperature $T_{g}$, i.e. $C_{G}\equiv G(T_{g})$. The constant $\varepsilon$ stems from the integration of $\alpha_T$ and from the dimensional prefactor in the power-law ansatz for $g(r)$. All the parameters in this expression are either fixed by the experimental/simulation protocol or can be found in the literature. The parameter $\lambda$ has to be extracted from $g(r)$ data, according to the protocol that we give in the Section VI.
%

\section{Analytical expression for the viscosity}
We can now use our model for $ G(T) $ to evaluate the activation energy $E(T)$ involved in restructuring the glassy cage and, hence, the viscosity $\eta(T)$ of the melts. Within the framework of the cooperative shear or elastic model of the glass transition~\cite{Eyring1943, Eyring1936, Dyre1998,Dyre2006}, the activation energy for local cooperative rearrangements is $E(T)=G V_{\text{c}}$. The characteristic atomic volume $V_{\text{c}}$ appearing here is accessible through the theoretical fitting of the viscosity data, although its value cannot be arbitrary and it must be representative of the atomic composition of the alloy and of the atomic sizes of its constituents.

Replacing the expression for $ E(T) $ in the Arrhenius relation given by the cooperative shear model of the glass transition, and using Eq.~\eqref{eq_th_shear} for $ G(T) $ inside $E(T)$, we obtain the following analytical expression for the viscosity, 
\begin{align}\label{thvisc}
\dfrac{\eta(T)}{\eta_{0}}=\exp{\left\{ \dfrac{V_c C_G}{k \, T}	\exp{\left[	(2+\lambda)	\alpha_T T_g \left(	1-\frac{T}{T_g}\right)	\right]}\right\}},
\end{align}
where $\eta_0$ is a normalisation constant set by the high-$T$ limit of $\eta$.

It is important to consider how the \textit{double-exponential} dependence of the viscosity on the temperature arises. The first exponential stems from the elastic activation described in the framework of the cooperative shear model, whereas the second exponential is due to the Debye-Gr\"uneisen thermal expansion rooted in lattice-dynamical
considerations of anharmonicity. This formula accounts for both anharmonicity, through $\alpha_{T}$, and for the repulsion steepness $\lambda$ (or softness $1/\lambda$).

\section{Estimating $\lambda$ from the radial distribution function of binary $ZrCu$ alloys}
We can now apply the tools introduced above to find an analytical connection between the interatomic interaction parameters and the $\eta(T)$ and $ G(T)$ of the melt up to $T_{g}$. To this end, we will apply our model to the binary system of $ZrCu$ alloys. The atomic-level structure of this system is studied by means of numerical simulations, and we present here also \textit{ad-hoc} experimental data for $\eta(T)$ and $g(r)$ for the special case of $Cu_{50} Zr_{50}$. Using our analytical model and the $g(r)$ data from simulations, we can build an analytical connection between $g(r)$ and $\eta(T)$, which makes use of the parameter $\lambda$ introduced above. The model will be tested against \textit{ad-hoc} experimental data of viscosity as a function of temperature for the $Cu_{50} Zr_{50}$ melt.

We also should notice that the $\lambda$ value is sensitive to the elemental composition and stoichiometry, but is less sensitive to temperature changes. This fact can be readily understood by considering that the interatomic potential results from the electronic structure of both valence and conduction electrons. In the temperature regime that we consider here and in the comparison with viscosity data below, it is quite unlikely that the electronic and ionic structure change with temperature, and this is reflected in the fact that the slope of the ascending flank of the the first peak of $g(r)$ does not change significantly in the range of $T$ under consideration. 
What changes with $T$, is the position of the first peak, of course, because the average distance between ions increases due to thermal expansion. 
For these reasons, in our comparison for the viscosity, below, we will use the value of $\lambda$ determined near $T_{g}$.

\subsection{Numerical simulations of $g(r)$ for $ZrCu$ alloys}
In order to extract the $\lambda$ parameter from sensible data, we carried out Molecular Dynamics (MD) simulations of the $Zr_{100 - x}$ system (where $x= 20, 35, 46, 50, 60, 65, 80$),  by employing an interatomic potential proposed by Duan et al. \cite{duan2005molecular2005}. This potential is a semi-empirical many-body potential developed in analogy to the tight-binding scheme in the second-moment approximation~\cite{cleri1993tight1993,rosato1989thermodynamical1989}. The equations of motion were integrated by using the Verlet algorithm with a time step of $5~\mathrm{fs}$. The configurations were prepared starting from a cubic cell box of $1.28\times 10^{5}$ atoms in the B2 structure and periodic boundary conditions in all directions. In each case, the positions of all atoms were redistributed randomly within the simulation cell and the resulting systems were first equilibrated at $300~\mathrm{K}$ in the isothermal-isobaric ensemble (NPT) for $100~\mathrm{ps}$ and subsequently heated up to $2000~\mathrm{K}$ for melting. After sufficient equilibration in the liquid state, the configurations were cooled down to $300~\mathrm{K}$ (always in the NPT ensemble) with a cooling rate of $10~\mathrm{K/ps}$, and they were finally equilibrated for $100~\mathrm{ps}$ (always in NPT). The structural changes of the system were studied by calculating the total $g(r)$ together with the Faber-Ziman partials every $100~\mathrm{K}$ upon cooling. In Fig.3 we show plots of the radial distribution function $g(r)$ for different compositions of the binary alloy. 

\subsection{Fitting protocol of $g(r)$ data}
Also shown in Fig. 3, are the fittings obtained using the power-law ansatz $g(r)\sim (r-\sigma)^{\lambda}$. In order to make quantitative fittings, the numerical coefficients in this ansatz need to be specified, so that we write
\begin{equation}
g(r)=g_{0}(r-\sigma+b)^{\lambda}.
\end{equation}
The numerical coefficients $g_{0}$ and $a$ take care of two important facts. Firstly, the $g(r)$ is dimensionless by its definition, hence the prefactor $g_{0}$ takes care of dimensionality and has dimensions of $\mathrm{\AA^{-\lambda}}$, if we express lengths in units of $\mathrm{\AA}$ in the above formula. Secondly, it is not realistic that $g(r)=0$, exactly, at $r=\sigma$. The probability density of two ions at that distance will be extremely small, but not identically zero. Hence, the parameter $b>0$ takes this fact into account. Finally, both parameters $g_{0}$ and $b$ control the value of $g(r)$ at the hard-core distance $\sigma = 2 a_{0}$, since $g(\sigma)=g_{0}b^{\lambda}$. Hence, the two parameters must satisfy the condition $g(\sigma)=g_{0} b^{\lambda}\lll 1$. 

In Ref.~\cite{Krausser}, we found that, in order to simultaneously fit shear modulus and viscosity data of different alloys, values of the exponent $\lambda$ in the order of magnitude $\lambda \sim 100$, are required. In turn, this constraint on the order of magnitude of $\lambda$, puts constraints on the orders of magnitude of the coefficients $g_{0}$ and $b$. Hence, for example, the only acceptable fitting of the $g(r)$ of the $Cu_{50} Zr_{50}$ alloy with $\lambda \sim 100$, can be achieved with $\lambda=80$. This value, in turn, constrains the values of the other coefficients to be $g_{0}=1 \times 10^{-28}$ and $b=2.16$ $\mathrm{\AA}$. 

Using the same values of the $g_{0}$ and $b$ coefficients also for the other compositions for which $g(r)$ was obtained from simulations, we produce the fittings reported in Fig. 3 for different values of stoichiometry.  

\begin{figure}
\includegraphics[width=0.94\columnwidth]{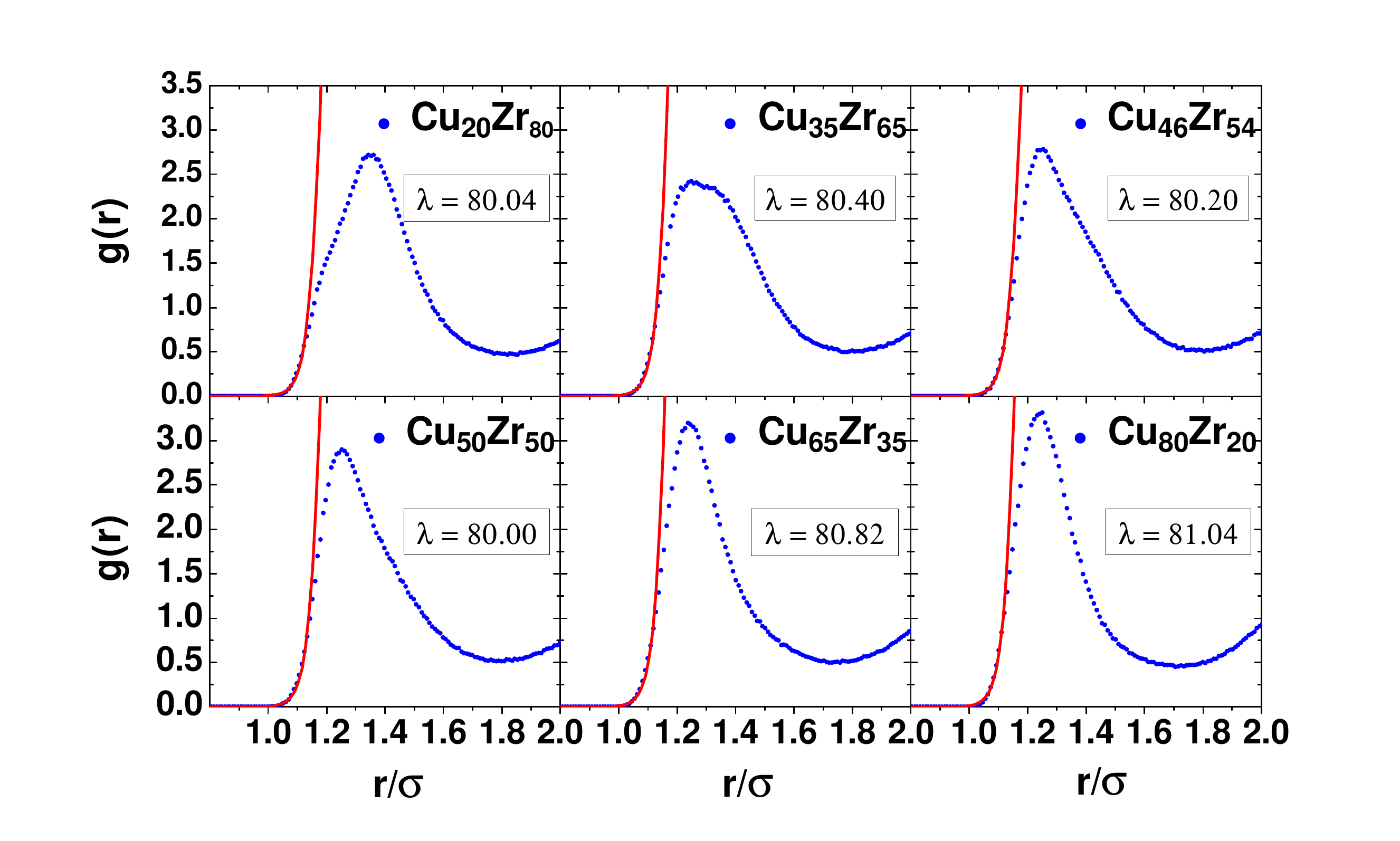}
\caption{(Color online)
Plots of $g(r)$ as obtained from numerical simulations (symbols) at $T=700$~$K$, together with the power-law fittings of the repulsive flank using Eq. (10) (solid lines). The values of the coefficients used for the fitting are $g_{0}=1.0 \times 10^{-28}$ and $b=2.16$ $\mathring{A}$ for all compositions, while the values of $\lambda$ reported in the figure panels and depend on the stoichiometry of the specific alloy, also reported in the figure panels. 
}
\label{fig3}
\end{figure}

\subsection{Dependence of the interatomic repulsion parameter $\lambda$ on the stoichiometry and the underlying electronic structure}
From the fittings we see that $\lambda$ is approximately constant for all alloys that are richer in $Zr$ up to the $Cu_{50} Zr_{50}$. As the alloy becomes richer in $Cu$, a trend becomes visible, where the $\lambda$ value increases as $Cu$ becomes the dominant component. This effect might be explained in terms of electronic structure of the ions, along the lines of~\cite{Krausser}. In particular, $Zr$-$Zr$ interactions may be effectively softer because of the pronounced $d$-wave character of the outer electron shells of $Zr$ which is a transition element, compared to the effectively more steeply repulsive $Cu$-$Cu$ interaction, dominated by the more pronounced $s$-wave character of the $Cu$ ions. Based on this picture, the $Zr$-$Cu$ interaction may be comparatively the most steeply repulsive, as the smaller $Cu$ atoms tend to nest in the corners, in
between the lobes of the $d$-wave structure of $Zr$.

\begin{figure}
\includegraphics[width=0.94\columnwidth]{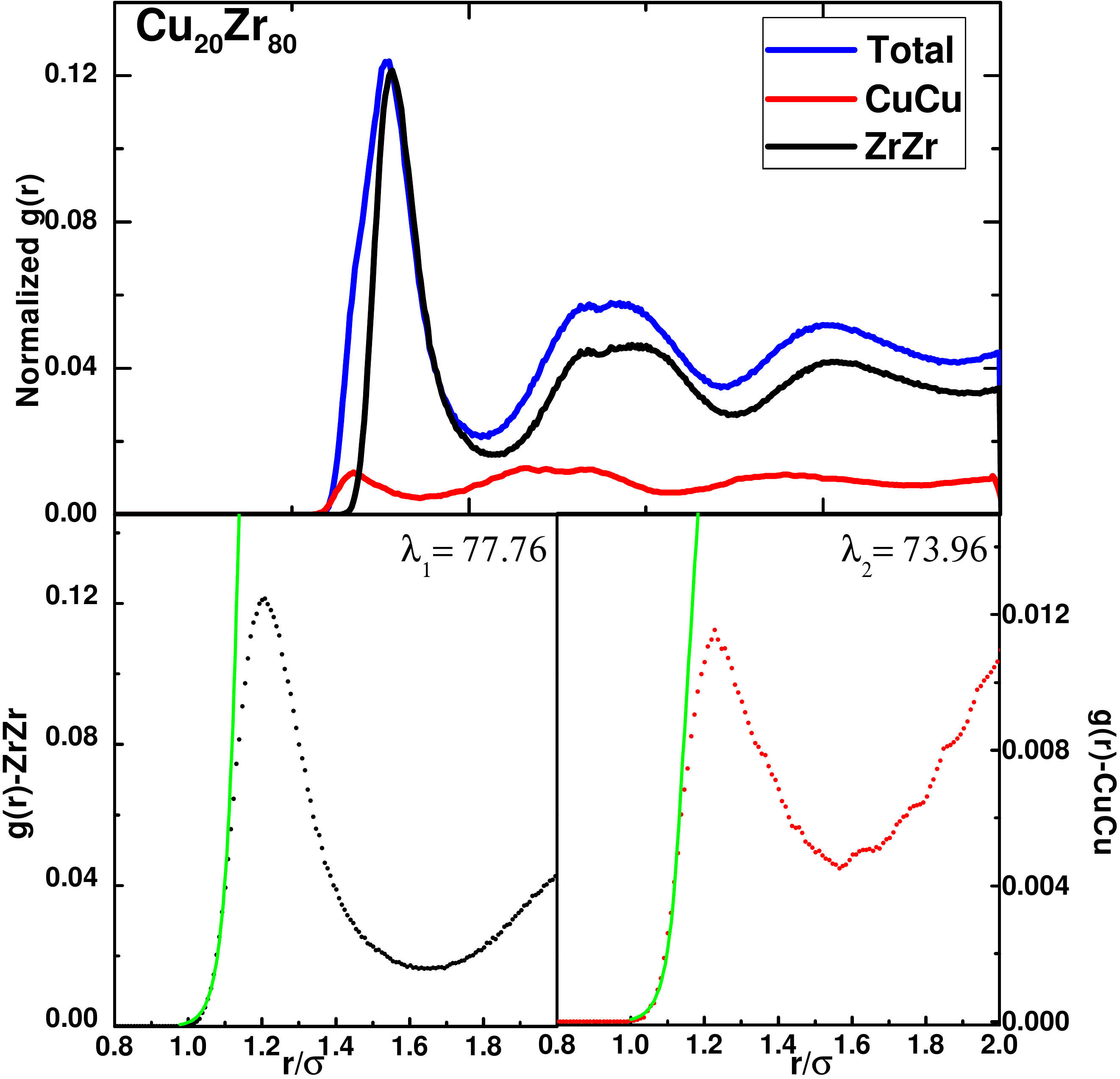}
\caption{(Color online)
Faber-Ziman partials of $g(r)$ for the $Cu_{20} Zr_{80}$ stoichiometry, together with their power-law fittings of the respective repulsive flanks using Eq. (10) (solid lines). The values of the coefficients used for the fitting are $g_{0}=1 \times 10^{-28}$ and $b=2.158$ $\mathring{A}$ for all compositions, while the values of $\lambda$ reported in the figure panels and depend on the stoichiometry of the specific alloy, also reported in the figure panels. 
}
\label{fig4}
\end{figure}

We also see from Fig. 3, that for the alloy richest in $Zr$, $Cu_{20} Zr_{80}$, two slopes in the repulsive ascending flank of the first peak are visible. In order to obtain more insight into this phenomenon, we have studied also the Faber-Ziman partials $g_{\alpha\beta}(r)$, where $\alpha,\beta=Zr,Cu$,  for this particular composition. The results are shown in Fig. 4. It is seen that the left-most slope is due to the $Cu$-$Cu$ contribution, which is steeper and shorter-ranged and thus contributes a higher value to the effective 
$\lambda$ which is larger than the partial contribution due to the $Zr$-$Zr$ contribution which is softer, as argued above because of the more pronounced $d$-wave character of the $Zr$ ion. The lower slope of the second flank is in fact due to the $Zr$-$Zr$ contribution. From this analysis, it appears that the short-range repulsive part of $g(r)$ is dominated by the $Zr$-$Cu$ contribution which is responsible for the $\lambda=80$ overall value, which varies only weakly with composition. This can be explained if one considers that smaller $Cu$ atoms tend to surround larger $Zr$ atoms and to nest into the corners of the $d$-wave outer shell of $Zr$.  We also note here that similar electronic interactions affect the slowing down of the dynamics in Zr-(Cu/Ni/Co) melts with Al additions via an enhanced short-range packing between the Al and late-transition metal species \cite{Cheng2009,Yuan2011,Yuan2015}.

\subsection{The interatomic repulsion parameter $\lambda$ is independent of $T$}
In order to explore a possible dependence of the interatomic repulsion steepness parameter $\lambda$ on the temperature $T$, we performed numerical simulations on the same alloy melts at different temperatures.
To this end, we have determined $g(r)$ from simulations of the $Cu_{50} Zr_{50}$ melt at $T=600$~$K$ and $T=800$~$K$, in addition to the $T=700$~$K$ data reported in Fig. 3. 

\begin{figure}
\includegraphics[width=0.94\columnwidth]{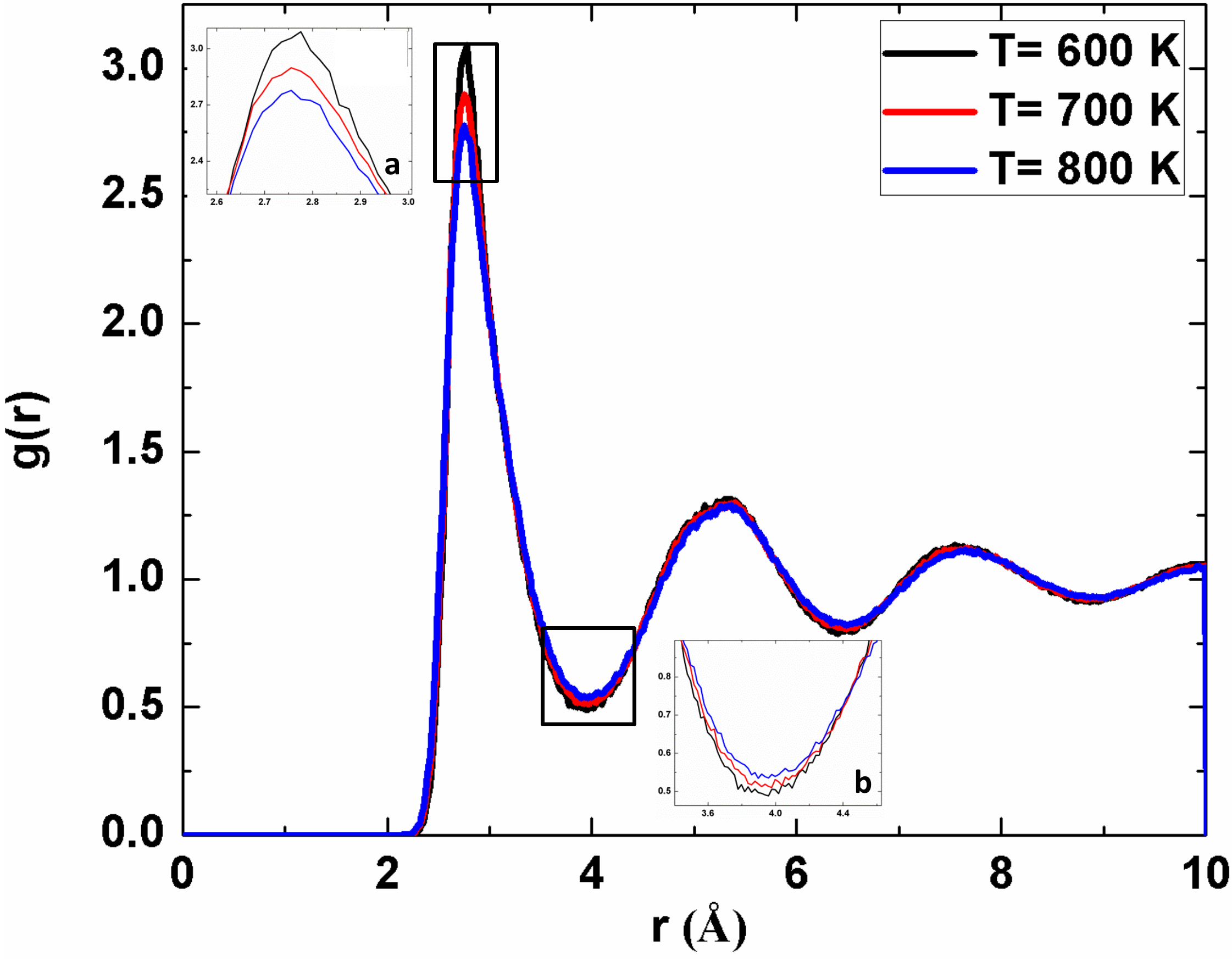}
\caption{(Color online)
Plot of the total simulated $g(r)$ for the Cu$_{50}$Zr$_{50}$ stoichiometry, at three different temperatures of the melt. 
}
\label{fig5}
\end{figure}

The comparison is shown in Fig. 5. As is evident, the only difference that comes from varying $T$ is the decrease in the height of the first peak of $g(r)$ upon increasing $T$. This effect is trivial and is attributed to the less pronounced smearing of atomic correlations as $T$ is increased~\cite{Stolpe2016}. Moreover, this observation clearly supports our main assumption that the connectivity changes with $T$, a manifestation of which is the fact that the parameter $Z$ in our model decreases upon increasing $T$. Thermal expansion greatly enhances this effect in experimental systems, as volume is not constant but is allowed to change following a change of temperature. 
It is also evident, that $T$ has no effect whatsoever on the slope of the repulsive left-hand-side flank of $g(r)$. The conclusion of this analysis is, therefore, that our repulsion steepness parameter $\lambda$, which is directly related to the slope of the left flank of $g(r)$, is independent of $T$. In the following steps of our model we will thus take $\lambda$ as independent of $T$.

\section{Comparison with experimental data of viscosity versus $T$}
We are now able to evaluate our analytical model linking the interatomic potential and the $g(r)$ with the macroscopic viscosity of the alloys upon approaching $T_{g}$ from the high-$T$ end. 
Since, unfortunately, experimental viscosity data over the wide range of $CuZr$ compositions modeled here are not readily available, we use experimental data for the $Cu_{50} Zr_{50}$ composition taken from \textit{ad-hoc} measurements as detailed in the Appendix. 

\begin{figure}[h]
	\centering
	\includegraphics[width=0.94\linewidth]{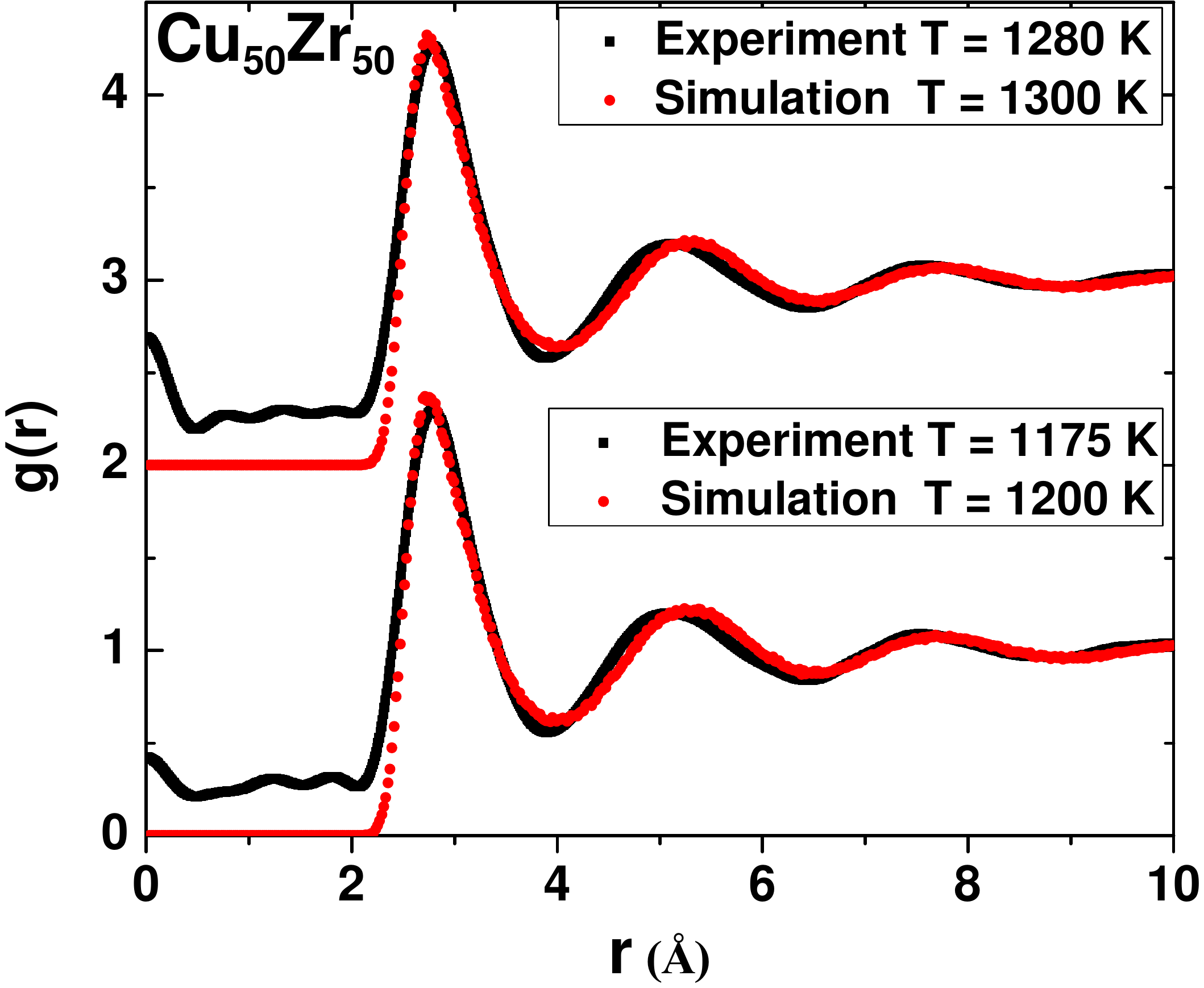}
	\caption{Comparison between RDFs of $ Cu_{\mathrm{50}}Zr_{\mathrm{50}} $ extracted from experiment and simulations. The dataset on top has been arbitrarily shifted for clarity. }
	\label{fig:fig6}
\end{figure}

The value of the interatomic repulsion steepness $\lambda= 80.0$ for this system was obtained from the fitting of $g(r)$ in Fig. 3. Using Eq. (9) it is therefore possible to obtain a one-parameter fitting of the experimental viscosity data for the $Cu_{50} Zr_{50}$ alloy as a function of $T$, as shown in Fig. 7. Furthermore, we recall from Eq. (8) that $C_{G}$ is defined as the value of the high-frequency shear modulus $G$ at $T=T_{g}$. For the $Cu_{50} Zr_{50}$ alloy, the value of $ C_G = 31.3~\mathrm{GPa} $ was obtained from experimental measurements in the literature~\cite{Johnson2006}.
The only adjustable parameter in Eq.(9) is thus the characteristic atomic volume $V_{c}$. From the fitting we obtain $V_{c}=0.0039~\mathrm{nm^3}$. 

In Fig. 7 (left panel) we also reported a fitting done with the VFT-type model of Ref. [4], which uses the concept of kinetic fragility $D^{*}$ to provide a more physical interpretation to the VFT relation (which normally has three adjustable parameters). The VFT-type relationship of Ref. [4] is given by $\eta(T)=a \exp[D^{*}T_{0}/(T-T_0)]$, where $a$ and $T_{0}$ are free parameters, while $D^{*}$ was adjusted to the experimental $Cu_{50} Zr_{50}$ data in Ref.[4] and interpreted as the kinetic fragility. In the fitting shown in Fig. 7, $T_{0}=578.6~K$ was used to obtain the best fitting, which is very far from the $T_{g}$ value of this system, $T_{g}=770~K$ as determined from our simulations and experiments for the $Cu_{50} Zr_{50}$ system. Therefore, the VFT-type fitting of Ref.[4] uses two free parameters, $D^{*}$ and $T_{0}$ (without considering the parameter $a$ which plays the same role as $\eta_{0}$ in our Eq.(9) and is set by the high-$T$ behaviour of viscosity). In contrast, our Eq. (9) has only one free parameter, $V_{c}$. In spite of having one adjustable parameter more than our model, the fitting of Ref.[4], provides a less accurate fitting in comparison, especially in the regime $T> 1200~K$.

The characteristic volume is defined within the framework of the cooperative shear model as $V_{c}=(2/3)(\Delta V)^{2}/V$ and was derived by Dyre \cite{Dyre2006} using linear elasticity for an expanding sphere. The quantity $\Delta V$ is the activation volume or the local volume change associated with a cooperative shear event and can be expressed as $\Delta V \approx \sqrt{480}V_c$. Furthermore, $V$ in these relations can be identified with the volume of a shear-transformation zone (STZ), and in Ref.~\cite{Krausser} it was found that, for various alloys, the approximate relationship $V\approx 320V_{c}$ holds.

The $V_{c}=0.0039$~$\mathrm{nm^{3}}$ value obtained here for $Cu_{50} Zr_{50}$ is on the same order of magnitude but smaller than the value obtained in our previous analysis~\cite{Krausser} for $Zr_{41.2}Ti_{13.8}Ni_{10}Cu_{12.5}Be_{22.5}$, for which a characteristic volume of $0.0085$~$\mathrm{nm^{3}}$ was found, which compares well with independent determinations on similar alloys~\cite{Krausser}. Moreover, for Cu$_{50}$Zr$_{50}$, we obtain an activation volume of $\Delta V \sim 100$~\AA$^{3}$, which is somewhat larger than $\Delta V = 84.3$~\AA$^3$, which was determined in MD simulations on a similar binary Cu$_{56}$Zr$_{44}$ alloy~\cite{Fan2014}.  
\begin{figure*} [!]
	{\includegraphics[width=0.95\columnwidth]{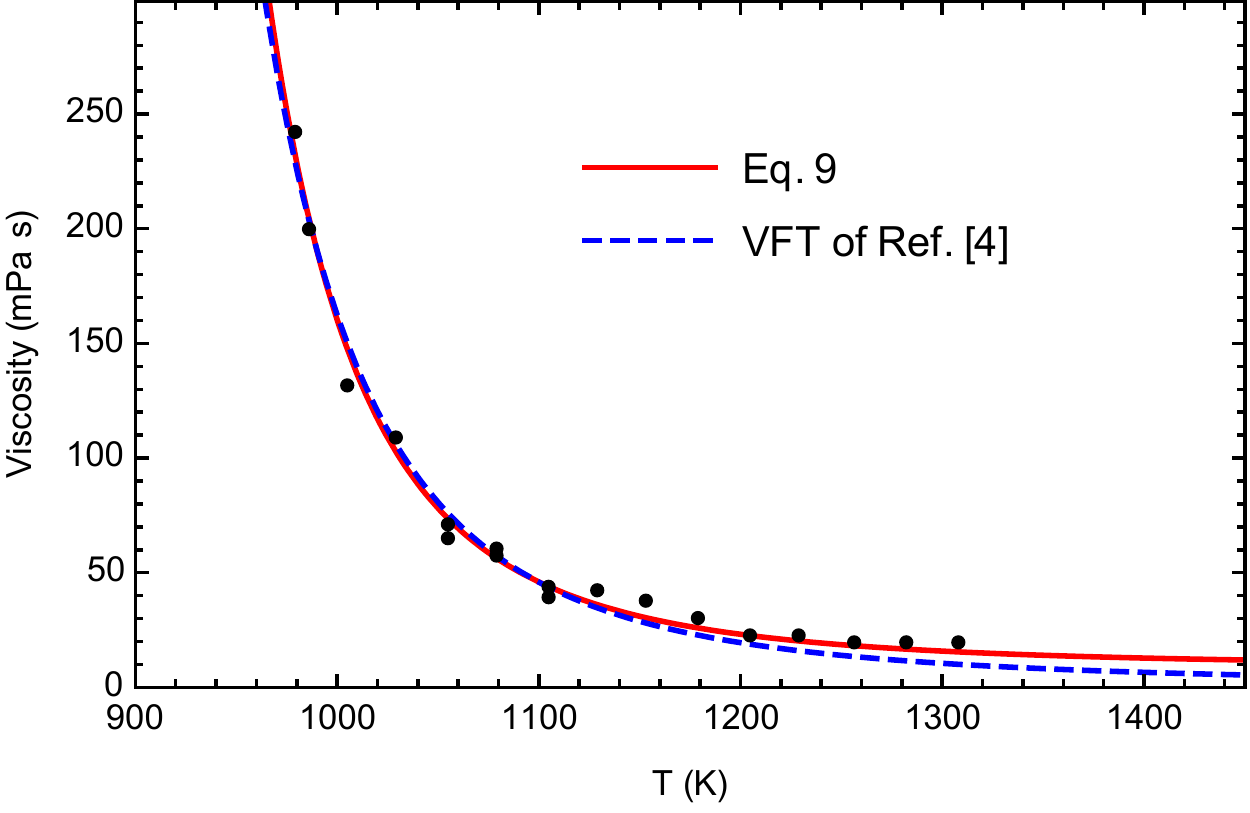}}
	{\includegraphics[width=0.937\columnwidth]{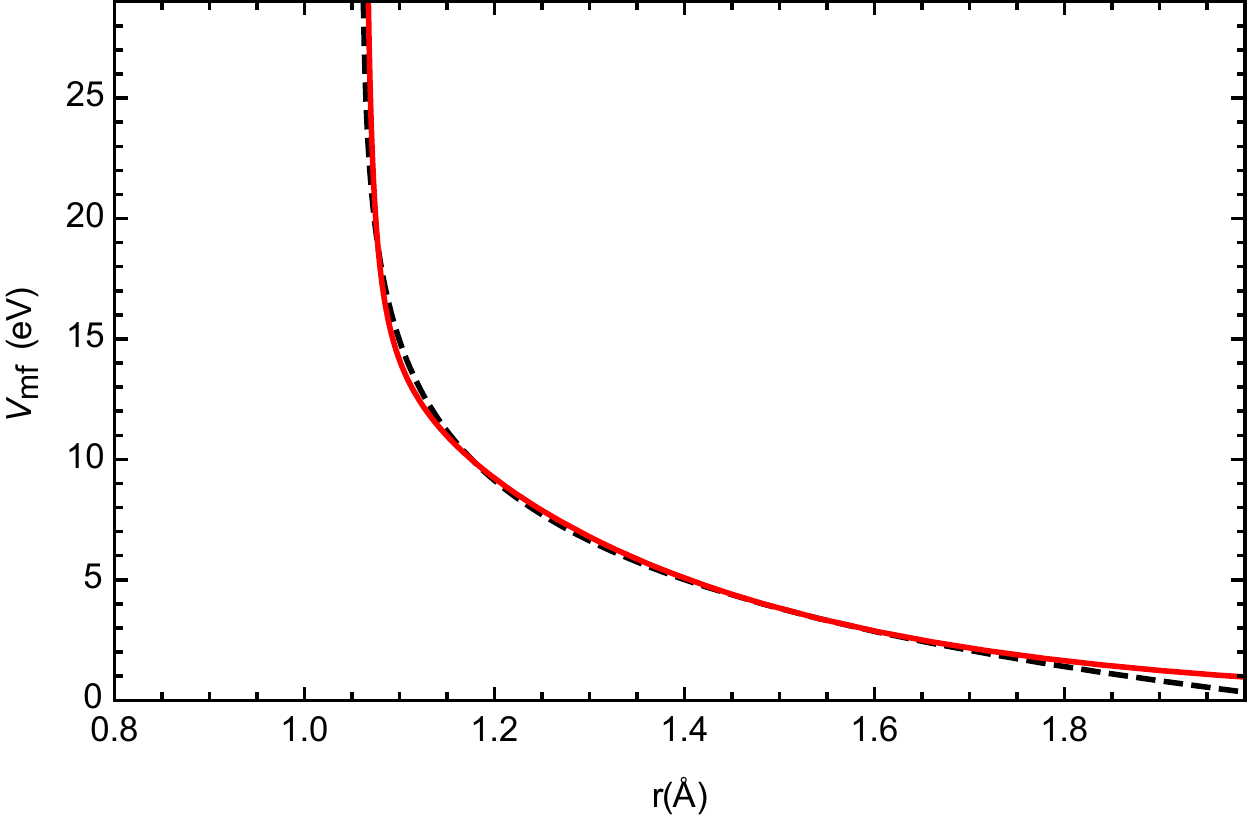}}
	\caption{(Color online) Calculated and experimental viscosity as a function of $T$ (left panel) for the alloy $Cu_{50} Zr_{50}$, and the underlying effective interatomic potential described by $\lambda=80$ (right panel). (a) Symbols are experimental data, the solid line is a one-parameter fit using Eq. (9), with the calculation parameters 
$\lambda=80$ determined from the $g(r)$ fitting and $C_{G}=31.3$~$\mathrm{GPa}$ from Ref.~\cite{Johnson2006}. $T_{g}=770~K$ is the glass transition temperature for the $Cu_{50} Zr_{50}$ system and is therefore not a fitting parameter. The only adjustable parameter is, hence, the characteristic atomic volume of the cooperative shear model $V_{c}=0.0046$~$\mathrm{nm^3}$, which turns out to be a meaningful value in comparison with previous estimates for similar alloys~\cite{Krausser}. The dashed line is a fitting made with the VFT model $\eta(T)=a\exp[D^{*}T_{0}/(T-T_0)]$ of Ref. [4]. Here three fitting parameters are used: $a=0.223$, $T_0=578.6$ and the kinetic fragility $D^{*}=4.8$. (b) The logarithmic potential of mean force from Boltzmann inversion of the repulsive flank of the $g(r)$ first peak, dashed line, and its fitting (solid line) using the Ashcroft-Born-Mayer interatomic potential, Eqs. (1)-(2), with parameters $A=0.121$~$\mathrm{eV}$, $q_{TF} = 1.7$~$\mathrm{\AA^{-1}}$, as in Ref.~\cite{Krausser}, $B=5.61$~$\mathrm{eV}$, $C= 2.8$~$\mathrm{\AA^{-1}}$, as in Ref.~\cite{Krausser}, and $\bar{\sigma}=1.35$~$\mathrm{\AA}$ from the average of ionic core sizes for $Cu_{50} Zr_{50}$.}
	\label{fig:6}
\end{figure*}

Finally, in Fig. 7 (right panel) we have also shown the underlying effective interatomic potential corresponding to the repulsion steepness value $\lambda=80$ obtained from the analysis of the $g(r)$, and its mapping to the Ashcroft-Born-Mayer potential in Eq.1. Some parameters used in the mapping are fixed and independent of the alloy composition, such as $q_{TF} = 1.7$~$\mathring{A}^{-1}$, from Ref.~\cite{Wang2004},  and $C= 2.8$~$\mathring{A}^{-1}$, perfectly within the range $1.89 - 4.72\mathring{A}^{-1}$ as discussed in Ref.~\cite{Krausser}. The parameters $A$ and $B$, instead, set the characteristic energy scales of the Ashcroft and Born-Mayer terms, respectively, in Eq. (1). The values that we found here for $Cu_{50} Zr_{50}$ are quite close to the values found in Ref.~\cite{Krausser} for other metallic glasses, which reflects the robustness of this approach.

\section{Conclusion}
In summary, we developed a protocol which takes the short-range repulsive part of $g(r)$ from simulation (or experimental) data as the input to qualitatively infer the viscosity of metallic alloys as a function of temperature upon vitrification, and their fragility.
Using a combination of simulation and experimental data for the case of the binary Cu-Zr alloys, it has been possible to produce a one-parameter theoretical fit of the viscosity as a function of temperature for the $Cu_{50} Zr_{50}$ system. 
Importantly, the steepness of the viscosity rise upon approaching the glass transition, is controlled by two averaged interaction parameters: the $\lambda$ interatomic-repulsion parameter introduced in Ref.~\cite{Krausser}, and the thermal expansion coefficient $\alpha_{T}$. 

This result confirms the picture according to which high thermal expansion coefficients favour fragile behaviour, alongside with steep short-ranged interatomic repulsion. The repulsion steepness is related to shallow changes at the level of nearest-neighbours upon decreasing $T$, whereas thermal expansion, associated with the long-range part of the interatomic potential, in metallic alloys is related to medium-range order effects and to the ability of the system to expel free volume at the level of the 3rd-4th coordination shells~\cite{Wei2015}.

In contrast to VFT-type relations for the viscosity as a function of $T$, our analytical theory is derived from the underlying atomic dynamics as shown in Ref.~\cite{Krausser}, and gives the viscosity as a double-exponential decreasing function of $T$. The outer exponential comes from the shoving model of the viscosity derived by Dyre, while the inner exponential is due to the exponential decrease of interatomic connectivity with increasing $T$ due to thermal expansion~\cite{Zaccone2013}. 
By estimating the repulsion parameter $\lambda$ from numerical simulations of $g(r)$ (validated at higher $T$ against experimental data), we have shown here how a quantitative fitting of viscosity as a function of $T$ can be obtained with just one fitting parameter, which is the characteristic atomic volume in the shoving model.
Also, all parameters involved in our viscosity expression have a clear microscopic physical meaning. 


\appendix
\section{Scaling between the connectivity $Z$ and the global interatomic repulsion steepness $\lambda$}
\label{connectivity}
When increasing $T$, the average spacing between atoms in the coordination shell becomes larger, and the probability of nearest neighbours leaving the connectivity shell increases. 
It is then possible to use the radial distribution function $g(r)$ to relate the change in atomic packing fraction $\phi$, due to an externally imposed change in temperature $T$, to the change in connectivity $Z$.
Following along the lines of Ref.~ \cite{Zaccone2013}, the connectivity can be written as usual as an integral over the $g(r)$, typically written as $Z  \approx \int_{0}^{r_{\mathit{max}}} r^2 g(r) dr$ up to a threshold which could be perhaps set conservatively as the maximum of $g(r)$. In amorphous systems, the peak of $g(r)$ increases upon increasing the packing fraction (and upon decreasing $T$ as is visible in Fig.5) due to many-body excluded-volume interactions, and this effect translates into a larger value of $Z$. A way to describe this effect, is to shift the upper limit of the integral which gives $Z$ in proportion to $\phi$, as $r_{\mathit{max}}=c\phi$, where $c$ is some proportionality constant, and one can write
\begin{equation}\label{integral}
Z  \approx \int_{0}^{c\phi} r^2 g(r) dr,
\end{equation}	
where $r$ represents the separation distance between two ions in the system. This relation correctly recovers the limit $Z\rightarrow 0$ in the limit $\phi\rightarrow 0$.

From Fig.5 it is also seen that the ascending part of the first peak of $g(r)$ happens within a narrow $r$-interval, which is quite close to the metallic diameter of the system, $\approx2.2 \mathring{A}$. At this very short separation between the ions, one can approximate the \textit{local} geometry as being effectively Cartesian, instead of spherical, which removes the metric factor from Eq. (A1). This is a customary simplification in dealing with small gaps between two spheres, and is widely in colloidal systems, where it is known as the Derjaguin approximation~\cite{Israelachvili2011}. Upon replacing our approximation for the ascending part of $g(r)$ Eq. (9) we thus obtain
\begin{equation}\label{integral}
Z  \approx \int_{0}^{c\phi} (r-\sigma+b)^{\lambda} dr.
\end{equation}	

Next, we make the standard change of variable in the integral $x=r-\sigma+b$, and the integral becomes
\begin{equation}\label{integral}
Z  \approx \int_{0}^{c\phi+q} x^{\lambda} dx,
\end{equation}
where $q$ is some coefficient. We then realize that the following approximation can be used 
\[\int_{0}^{c\phi+q} x^{\lambda} dx = \int_{0}^{q} x^{\lambda} dx + \int_{q}^{q+c\phi} x^{\lambda} dx \sim (c\phi)^{1+\lambda}.\]

Hence we obtain the key asymptotic scaling
\begin{equation}
Z\sim \phi^{1+\lambda},
\end{equation}
used in our model to link the $g(r)$ with the high-frequency shear modulus $G$ and then with the viscosity $\eta$.

\section{Experimental measurement of $g(r)$ and viscosity for the $Cu_{50} Zr_{50}$ system}

The $g(r)$ of the $Cu_{50}Zr_{50}$ melt was determined previously in combined neutron and synchrotron X-ray diffraction experiments, described in detail in Ref.~\cite{Holland-Moritz2012}
	. From the total neutron and X-ray static structure factors $S(q)$, two of the three Bhatia-Thornton partial structure factors~\cite{Bhatia1970}, $S_{nn}(q)$ and $S_{nc}(q)$, were able to be determined with good precision. The corresponding $g_{nn}(r)$ and $g_{nc}(r)$ were calculated by Fourier transformation of $S_{nn}(q)$ and $S_{nc}(q)$, respectively.

The viscosity values were determined in both the equilibrium and undercooled states of the Cu$_{50}$Zr$_{50}$ melt at the Institute for Materials Physics in Space of the German Aerospace Center (DLR) in Cologne, Germany. An electrostatic levitation (ESL) apparatus under ultra-high vacuum conditions of $\approx 10^{-8}$\,mbar was used to levitate an electrically charged sample with a mass of 50\,mg. Heating was achieved via two 25 W IR lasers and the temperature measured contact-free by a pyrometer directed at the sample side. A high-speed camera allowed determination of the vertical sample radius $R_z(t)$ as a function of time $t$. The levitated sample was brought in to the liquid state and heated to an initial temperature of 1308\,K, which is 100\,K higher than the liquids temperature $T_{liq} = 1208$\,K, taken from the published phase diagram~
		\cite{ZrCuPhase1990}
	. Measurements of the viscosity were carried out isothermally during cooling using the oscillating droplet method 
		\cite{Brillo2011}
	. The viscosity was determined at each temperature by measuring the decay of surface oscillations induced by a sinusoidal electric field described by
\begin{equation}
R_z(t) = R_0 + A\exp(-t/\tau_0)\sin(\omega t + \delta_0),
\end{equation}
where $R_0$ is the quiescent sample radius, $A$ the oscillation amplitude, $\tau_0$ the decay time constant, $\omega$ the frequency and $\delta_0$ the phase shift. Using Lamb's law~
		\cite{Lamb1993}
	, the viscosity was calculated as
\begin{equation}
\eta = \frac{\rho R_0^2}{5\tau_0},
\end{equation}
where $\rho$ is the macroscopic density of the droplet, also determined in these ESL experiments combined with video diagnostic techniques~
		\cite{Brillo2011,Yang2014}. \\

\begin{acknowledgments}
We are especially thankful to A. Meyer, F. Yang and D. Holland-Moritz for providing the experimental $g(r)$ data and access to the ESL apparatus for the viscosity measurements. \newline
The support of the EU through VitrimetTech ITN network FP7-PEOPLE-2013-ITN-607080 is fully acknowledged.

\end{acknowledgments}


\begin{thebibliography}{99}
	
\bibitem{Royall2008} Royall C P, Williams S R, Ohtsuka T and Tanaka H,
\textit{Direct observation of a local structural mechanism for dynamical arrest}, 2008 Nature Materials \textbf{7} 556

\bibitem{Milkus2016} Milkus R and Zaccone A, \textit{Local inversion-symmetry breaking controls the boson peak in glasses and crystals}s, 2016 Physical Review B \textbf{93} 094204

\bibitem{Ketov2015} Ketov S V et al.,\textit{ Rejuvenation of metallic glasses by non-affine thermal strain}, 2015 Nature \textbf{524} 200 

\bibitem{Berthier} Berthier L and Biroli G \textit{Theoretical perspective on the glass transition and amorphous materials}, 2011 \textbf{83} 587

\bibitem{Mauro2014}	Mauro N A, Blodgett M, Johnson M L, Vogt A J and Kelton K F, \textit{A structural signature of liquid fragility}, 2014 Nat Commun \textbf{5} 1

\bibitem{Weeks1971}	 Weeks J D, Chandler D and Andersen H A, \textit{Role of Repulsive Forces in Determining the Equilibrium Structure of Simple Liquids}, 1971 Journal of Chemical Physics \textbf{54} 5237


\bibitem{Mattson2009} Mattsson J, Wyss H M, Fernandez-Nieves A, Miyazaki K, Hu Z, Reichman D R and Weitz D A, \textit{Soft Colloids Make Strong Glasses}, 2009 Nature \textbf{462} 83 

\bibitem{Krausser} Krausser J,  Samwer K, Zaccone A, \textit{Interatomic repulsion softness controls the fragility of supercooled metallic melts}, 2015 Proc. Natl. Acad. Sci. USA \textbf{112} 13762

\bibitem{Huang2011}	Huang B, Bai H Y and Wang W H, \textit{Unique properties of CuZrAl bulk metallic glasses induced by microalloying}, 2011 Journal of Applied Physics \textbf{110} 123522
	
\bibitem{Wei2015}Wei S, Stolpe M, Gross O, Evenson Z, Gallino I,  Hembree W, Bednarcik J, Busch R, \textit{Linking structure to fragility in bulk metallic glass-forming liquids}, 2015 Applied physics letters \textbf{106} 181901

\bibitem{Jaiswal} Jaiswal A, Egami T, Kelton K F, Schweizer K S and  Zhang Y, \textit{Correlation between fragility an d Arrhenius crossover phenomenon in metallic, molecular, and network liquids}, arXiv:1604.08920v1  [cond-mat.soft]. 
 
\bibitem{Wang2015}
Wang Y-J, Zhang M, Liu L, Ogata S, Dai L H,
\textit{Universal enthalpy-entropy compensation rule for the deformation of metallic glasses}, 2015 Phys. Rev. E \textbf{92} 174118
 
 
\bibitem{Hu2015} Hu Y C, Li F X, Li M Z, Bai H Y and Wang W H, \textit{Five-fold symmetry as indicator of dynamic arrest in metallic glass-forming liquids}, 2015 Nature Commun. \textbf{6} 8310 

\bibitem{Lagogianni2009} Lagogianni A E, Almyras G, Lekka C E, Papageorgiou D G and Evangelakis G A, \textit{Structural characteristics of $Cu_{x}$ $Zr_{100 - x}$ metallic glasses by Molecular Dynamics Simulations}, 2009 Journal of Alloys and Compounds \textbf{483} 658

\bibitem{Bokas2013} Bokas G B, Lagogianni A E, Almyras G and Evangelakis G A, \textit{On the role of Icosahedral-like clusters in the solidification and the mechanical response of Cu-Zr metallic glasses by Molecular Dynamics simulations and Density Functional Theory computations}, 2013 Intermetallics \textbf{43} 138 

\bibitem{Almyras2011}  Almyras G, Papageorgiou D G, Lekka C E, Mattern N, Eckert J and Evangelakis G A  \textit{Atomic cluster arrangements in reverse monte carlo and molecular dynamics structural models of binary Cu-Zr metallic glasses}, 2011 Intermetallics \textbf{19} 657

\bibitem{Antonowicz2012} Antonowicz J, Pietnoczkaa A, Drobiazga T, Almyras G A, Papageorgiouc D G and Evangelakis G A, \textit{Icosahedral order in Cu-Zr amorphous alloys studied by means of X-ray absorption fine structure and molecular dynamics simulations}, 2012 Philosophical Magazine \textbf{92} 1865



\bibitem{Johnson2007} Johnson W L, Demetriou M D, Harmon J S, Lind M L and Samwer K, \textit{Rheology and ultrasonic properties of metallic glass-forming liquids: A potential energy landscape perspective}, 2007 MRS bulletin \textbf{32} 644

\bibitem{Jensen2014}Jensen K E, Weitz D A and Spaepen F, \textit{Local shear transformations in deformed and quiescent hard-sphere colloidal glasse}s, 2014 Physical Review E \textbf{90} 042305

\bibitem{Faber1972} Faber TE, \textit{Introduction to the Theory of Liquid Metals}, 1972 Cambridge Univ Press, Cambridge, UK

\bibitem{Shannon1976} Shannon RD,  \textit{Revised effective ionic radii and systematic studies of interatomic distances in halides and chalcogenides}, 1976 Acta Crystallogr A \textbf{32} 751


\bibitem{Wang2004} Wang W, Dong C, Shek C  \textit{Bulk metallic glasses}, 2004 Mater Sci Eng R Rep \textbf{44} 45
 
\bibitem{Nikulin1970} Gaydaenko V and  Nikulin V, \textit{ Born-Mayer interatomic potential for atoms with Z =2 to Z =36}, 1970 Chem Phys Lett \textbf{7} 360




\bibitem{Hafner1987} Hafner J, \textit{ From Hamiltonians to Phase Diagrams}, 1987 Springer Series in Solid-State
Sciences (Springer, Berlin) 70.

\bibitem{Abrahamson1969} Abrahamson AA,  \textit{Born-Mayer-type interatomic potential for neutral groundstate atoms with Z =2 to Z =105}, 1969 Phys Rev \textbf{178} 76

\bibitem{Zwanzig1965} Zwanzig R and Mountain R D, \textit{ High-Frequency Elastic Moduli of Simple Fluids}, 1965 The Journal of Chemical Physics \textbf{43} 4464

\bibitem{Lemaitre2006}
Lemaitre A and Maloney C, \textit{Sum Rules for the Quasi-Static and Visco-Elastic
	Response of Disordered Solids at Zero Temperature}, 2006 J. Stat. Phys. \textbf{123} 415


\bibitem{Zaccone2011a}
Zaccone  A \& Scossa-Romano E, \textit{Approximate analytical description of the nonaffine response
  of amorphous solids}, 2011 Phys. Rev. B {\bf 83} 184205.

\bibitem{Zaccone2011b}
Zaccone A, Blundell, J R,  \& Terentjev E M,\textit{ Network disorder and nonaffine deformations in marginal
  solids}, 2011 Phys. Rev. B {\bf 84} 174119

\bibitem{Zaccone2013}
Zaccone A \& Terentjev E M, \textit{Disorder-Assisted Melting and the Glass Transition in Amorphous Solids}, 2013 Phys. Rev. Lett {\bf 110} 178002

\bibitem{Rizzi2016}
Rizzi, L. G., S. Auer, and D. A. Head. \textit{Importance of non-affine viscoelastic response in disordered fibre networks}, 2016 Soft Matter \textbf{12} 4332.


\bibitem{Zaccone2014}
Zaccone A, Schall P  \& Terentjev E M,\textit{ Microscopic origin of nonlinear nonaffine deformation in bulk metallic glasses}, 2014 Phys. Rev. B {\bf 90} 140203

\bibitem{Wyart}
Wyart, M, \textit{On the rigidity of amorphous solids}, 2005 
\newblock {\em Ann. Phys. (Fr.)} {\bf 30}, 1.


\bibitem{Egami2002}
Egami T, \textit{Nano-glass Mechanism of Bulk Metallic Glass Formation}, 2002 Materials Transactions \textbf{43} 510

\bibitem{Khonik}   
Makarov AS, Khonik VA, Mitrofanov Yu P, Tsyplakov AN, \textit{Prediction of the annealing effect on room-temperature shear modulus of a metallic glass}, 2016 Intermetallics 
\textbf{69} 10.


\bibitem{Eyring1943}
Tobolsky A, Powell R E \& Eyring H, \textit{Elastic-viscous properties of matter}, 1943  Front. Chem. {\bf 1} 125



\bibitem{Eyring1936}
Eyring, H.
\textit{{Viscosity, Plasticity, and Diffusion as Examples of Absolute
  Reaction Rates}}, 1936
 J. Chem. Phys. {\bf 4} 283

\bibitem{Dyre1998}
Dyre J C,
  \textit{{Source of non-Arrhenius average relaxation time in
  glass-forming liquids}}, 1998 J. Non. Cryst. Solids {\bf 142} 235

\bibitem{Dyre2006}
Dyre  J C,
\newblock \textit{Colloquium: The glass transition and elastic models of
  glass-forming liquids}, 2006 Rev. Mod. Phys. {\bf 78} 953




\bibitem{duan2005molecular2005}
Duan G, Xu D, Zhang Q, Zhang G, Cagin T, Johnson W and Goddard~III W, \textit{Molecular dynamics study of the binary $Cu_{46}Zr_{54}$ metallic glass motivated by experiments: Glass formation and atomic-level structure}, 2005 Phys. Rev. B \textbf{71} 224208 
 
\bibitem{cleri1993tight1993}
Cleri F and Rosato V, \textit{Tight-binding potentials for transition metals and alloys}, 1993 Phys. Rev. B \textbf{48} 22

\bibitem{rosato1989thermodynamical1989}
Rosato V, Guillope M and Legrand B, \textit{ Thermodynamical and structural properties of fcc transition metals using a simple tight-binding model}, 1989 Philos. Mag. A \textbf{59} 321






\bibitem{Cheng2009}
Cheng Y, Ma E and Sheng H, \textit{Atomic Level Structure in Multicomponent Bulk Metallic Glass}, 2009 Phys. Rev. Lett. \textbf{102} 245501 

\bibitem{Yuan2011}
Yuan C C, Xiang J F, Xi X K and Wang W H, \textit{NMR Signature of Evolution of Ductile-to-Brittle Transition in Bulk Metallic Glasses}, 2011 Phys. Rev. Lett. \textbf{107} 236403

\bibitem{Yuan2015}
Yuan C C, Yang F, Kargl F, Holland-Moritz D, Simeoni G G and Meyer A, \textit{Atomic dynamics in Zr-(Co,Ni)-Al metallic glass-forming liquids}, 2015 Phys. Rev. B \textbf{91}, 214203 






\bibitem{Stolpe2016}
Stolpe M, Jonas I, Wei S, Evenson Z, Hembree W, Yang F, Meyer Aand Busch R, \textit{Structural changes during a liquid-liquid transition in the deeply undercooled Zr$_{58.5}$Cu$_{15.6}$Ni$_{12.8}$Al$_{10.3}$Nb$_{2.8}$ bulk metallic glass forming melt}, 2016 Phys. Rev. B \textbf{93} 014201 

\bibitem{Johnson2006}
G.J. Fan et al., \textit{Thermophysical and elastic properties of Cu50Zr50 and
	$(Cu50Zr50)_{95}Al_{5}$ bulk-metallic-glass-forming alloys}, 2006 Appl. Phys. Lett. \textbf{89 }241917


\bibitem{Fan2014}
Fan Y, Iwashita T and Egami T, \textit{How thermally activated deformation starts in metallic glass}, 2014 Nat. Commun. \textbf{5}


\bibitem{Israelachvili2011}
Israelachvili J, \textit{Intermolecular and surface forces}, 2011 Elsevier Amsterdam 2011 


\bibitem{Holland-Moritz2012}
Holland-Moritz D, Yang F,  Kordel T, Klein S, Kargl F, Gegner J, Hansen T, Bednarcik J, Kaban I, Shuleshova O, Mattern N and Meyer A, \textit{Does an icosahedral short-range order prevail in glass-forming Zr-Cu melts?}, 2012 EPL \textbf{100} 56002

\bibitem{Bhatia1970}
Bhatia A B and Thornton D E, \textit{Structural Aspects of the Electrical Resistivity of Binary Alloys}, 1970 Phys. Rev. B \textbf{2} 3004 

\bibitem{ZrCuPhase1990}
Arias D and Abriata J P, \textit{Bulletin of Alloy Phase Diagrams}, 1993 \textbf{11}, 452

\bibitem{Brillo2011}
Brillo J, Pommrich A and Meyer A, \textit{Relation between Self-Diffusion and Viscosity in Dense Liquids: New Experimental Results from Electrostatic Levitation}, 2011 Phys. Rev. Lett. \textbf{107} 165902



\bibitem{Lamb1993}
Lamb H, \textit{Hydrodynamics, 6th ed.}, 1993 Cambridge University Press Cambridge UK

\bibitem{Yang2014}
Yang F, Holland-Moritz D, Gegner J, Heintzmann P, Kargl F, Yuan C C, Simeoni G G and Meyer A, \textit{Atomic dynamics in binary Zr-Cu liquids}, 2014 EPL \textbf{107} 46001




%
%
%
%




\end{thebibliography}
\end{document}